\documentclass[global,twocolumn]{svjour}
\usepackage{graphics}
\journalname{appliedphysicsB}
\begin{document}
\title{Characterization and control of phase fluctuations in elongated Bose-Einstein condensates}
\author{%\inst{2}
H. Kreutzmann\inst{1} \and A. Sanpera\inst{1} \and L. Santos\inst{1}\and 
M. Lewenstein\inst{1}
\newline
D. Hellweg\inst{2} \and L. Cacciapuoti\inst{2} \and M. Kottke\inst{2} \and T.
Schulte\inst{2} \and K. Sengstock\inst{3} \and J. J. Arlt\inst{2} \and and W.
Ertmer\inst{2}
}                     % Do not remove
%
% \offprints{}          % Insert a name or remove this line
%
\institute{Institut f\"ur Theoretische Physik, Universit\"at Hannover, Appelstra\ss e 2, 30167 Hannover, Germany
\and
Institut f\"ur Quantenoptik, Universit\"at Hannover, Welfengarten 1, 30167 Hannover, Germany
\and 
Institut f\"ur Laserphysik, Universit\"at Hamburg, Jungiusstra\ss e 9, 20355 Hamburg, Germany}
\date{Received: date / Revised version: date}
\maketitle
\begin{abstract}

% Abstract

Quasi one dimensional Bose-Einstein condensates (BECs) in elongated
traps exhibit significant phase fluctuations even at very low
temperatures. 
We present recent experimental results on the dynamic transformation
of phase fluctuations into density modulations during time-of-flight and
show the excellent quantitative agreement with the theoretical prediction.
In addition we confirm that under our experimental conditions, in the magnetic trap 
density modulations are strongly suppressed even when the phase
fluctuates. The paper also discusses our theoretical 
results on control of the condensate phase
by employing a time-dependent perturbation.
Our results set important limitations on
future applications of BEC in precision atom interferometry and atom
optics, but at the same time suggest pathways to overcome these limitations.

\end{abstract}
\section{Introduction}
\label{intro}
Since the first experimental observation of Bose-Einstein condensates~\cite{bec}
a considerable effort has been devoted to studies
of low dimensional ultra-cold trapped gases~\cite{low1d}. There are several
fundamental reasons for the interest in these systems.
Firstly, atom-atom
interactions can be modified in low dimensions~\cite{gorarev}.
Secondly, the physics in low dimensional systems can be
remarkably different from the situation in three dimensions.
For example, a one dimensional condensate in the limit of 
ultra-low densities should behave as a strongly correlated
system of impenetrable bosons, the so-called Tonks gas~\cite{tonks}.
The third, and perhaps the most important reason to study low temperature
1D condensates, is the particular potential that such 
condensates offer for applications. Proposals  of atom optics and
precision atom interferometry with coherent
matter waves rely on placing the condensate in a 1D waveguide.
Such a waveguide can be formed in appropriately designed traps~\cite{hanwaveguide} or on
the surface of an atom chip~\cite{waveguides,ChipBEC}. One could then use 
the BEC on such a chip to perform
beam splitting and, finally,  beam recombination in a waveguide.

However, it has been shown that BEC in quasi 1D
geometries exhibit significant phase
fluctuations~\cite{Shlyapnikov1D,Shlyapnikov3D}.
Even at very low temperatures (of the order
of fractions of the critical temperature $T_{\mbox{c}}$)
the coherence length of such a phase-fluctuating BEC may be
shorter than the size of the condensate.
The condensate then exhibits a spatially varying phase pattern and is
commonly called a \textit{quasicondensate}.
In the Thomas-Fermi regime~\cite{Dalfovo}, that is, when the nonlinear
mean field energy is much larger than the kinetic energy, 
density fluctuations are suppressed due to their energetic
cost. That is not the case, however, for the phase fluctuations. 
The prediction of
phase fluctuations~\cite{Shlyapnikov1D} was
experimentally confirmed in~\cite{Dettmer,Hellweg}. 

The results of our experiments set important limitations on the
future applications of BEC in interferometry  which employ 1D waveguides.
It is, therefore, particularly important 
to study and characterize phase fluctuations
in elongated BECs and, even more, to develop methods that allow to control
the phase of these condensate.

The first part of this paper concerns the experimental investigation
of two aspects of phase-fluctuating BECs.
First the dynamical behavior during a time-of-flight after release from a strongly elongated trap is examined. 
Very good quantitative agreement between theory and experiment is established for these measurements.
A second set of measurements is devoted to the
investigation of the density of a phase-fluctuating trapped condensate.
We confirm that the density fluctuations are strongly suppressed and show
that the
second order correlation function $g^{(2)}(0)$ is
largely independent of the amount of phase
fluctuation in our experimental regime.

The second part of the paper deals with theoretical approaches to
overcome the effects of phase fluctuations. 
We propose to use the technique of parametric 
resonance to control the overall phase of the condensate.
It should be noted that parametric resonance has already been 
discussed in the context of BEC~\cite{garciari1,kevrekidis1,staliunas}, but 
those studies were aimed at employing the
parametric resonance to create patterns and textures 
in the density of trapped BECs. 
Clearly, for the reasons mentioned above, 
such a goal is not easily accomplished in highly
nonlinear systems for which fluctuations of the mean field 
energy are costly. We, however, aim at
controlling the phase of a quasi 1D condensate by applying
the standard technique of modulating
the trap frequency~\cite{modulationtrap}.
Thus, rather than to excite 
density modulations of the condensate, we intend  
to engineer its phase, 
such that some modes contributing to the phase fluctuations
can be suppressed or enhanced 
without destroying the condensate.

The paper is organized as follows: 
In Section 2 we briefly review the physics of
phase fluctuations in elongated traps, 
and recall the expressions describing the
transformation of phase fluctuations into density modulations~\cite{Dettmer}.
In Section 3 we present recent experimental results which 
expand the observations reported
in our earlier work. These results show that a full quantitative
characterization of phase fluctuations can be achieved
and that density modulations are suppressed even when
phase fluctuations are present in a trapped condensate.
In Section 4, we describe the method of 
employing parametric resonance to
enhance or suppress selected modes of the phase fluctuations. 
We present model calculations based on
numerical solutions of the 1D Gross-Pitaevskii (GP)
equation which governs the dynamics of the 
condensate.

\section{Description of the phase fluctuations}
\label{theory}

For BECs in 3D trapping geometries
the fluctuations of density and phase are 
only important in a narrow  temperature range near the BEC 
transition temperature $T_{\mbox{c}}$ 
(critical fluctuations). 
Outside this region, the fluctuations are suppressed and the condensate 
is phase and density coherent. In reduced dimensions, the situation
can be rather different
(see~\cite{Shlyapnikov1D,Shlyapnikov2D} and refs. therein):
the axial phase fluctuations can manifest 
themselves even at temperatures far below $T_{\mbox{c}}$.

In the following we consider a cylindrically symmetric
condensate in the Thomas-Fermi regime,
where the mean field (repulsive) interparticle
interaction greatly exceeds the radial ($\hbar \omega_{\rho}$) and axial
($\hbar \omega_x$) trap energies. At 
$T=0$ its density profile has the well-known shape $n_0(\rho,x)=n_{0\mbox{m}}
(1-\rho^2/R_{\mbox{\tiny TF}}^2-x^2/L_{\mbox{\tiny TF}}^2)$,
where $n_{0\mbox{m}}=\mu/g$ is the maximum
condensate density,  $\mu$ is the chemical potential,
$g=4\pi\hbar^2a/m$ is the interaction coupling constant, 
$m$ is the atomic mass, and $a>0$ the scattering length.
Under the condition $\omega_{\rho}\gg\omega_x$, the radial size of the
condensate, given by the Thomas-Fermi radius 
$R_{\mbox{\tiny TF }}=(2\mu/m\omega_{\rho}^2)^{1/2}$, is much smaller  than the axial
size, given by the corresponding Thomas-Fermi length
$L_{\mbox{\tiny TF }}=(2\mu/m\omega_x^2)^{1/2}$.   

The phase fluctuations can be described by solving the 
Bogoliubov-de Gennes equations
describing elementary excitations of the condensate. 
The quantum field annihilation operator of atoms can be written as
$\hat\psi({\bf r})=\sqrt{n_0({\bf r})}\exp(i\hat\phi({\bf r}))$,
where $\hat\phi({\bf r})$ is the operator of
the phase, and the density fluctuations have already been neglected
following the arguments discussed above. 
The single-particle correlation function is then
expressed\\
through the mean square fluctuations of the phase (see e.g.~\cite{Popov}):
\begin{equation}  
\label{opdm} 
\!\langle\hat\psi^{\dagger}({\bf r})\hat\psi({\bf
r}')\rangle\!=\!\sqrt{n_0({\bf r})n_0({\bf r}')} 
\exp\{-\langle[\delta\hat\phi({\bf r},{\bf r}')]^2\rangle/2\},\!
\end{equation} 
with $\delta\hat\phi({\bf r},{\bf r}')=\hat\phi({\bf r})-\hat\phi({\bf
r}')$. The operator $\hat\phi({\bf r})$ is given by (see e.g.~\cite{Shev})
\begin{equation}
\label{operphi}
\hat\phi({\bf r})=[4n_0({\bf r})]^{-1/2}\sum_{j}f_j^{+}({\bf r})\hat a_j 
+\mbox{h.c.},
\end{equation}
where $ \hat{a}_j$ is the annihilation operator of the excitation with
quantum number $j$ and energy $\epsilon_j$, $f_j^{+}= u_j +
v_j$, and the excitation functions $u_j,v_j$ are determined by the
Bogoliubov-de Gennes equations
in the Thomas-Fermi limit. 

The ``low-energy'' axial excitations (with energies\\
$\epsilon_{j}<\hbar\omega_{\rho}$)
have wavelengths larger than $R_{\mbox{\tiny TF}}$ and exhibit a pronounced 1D behavior. Hence,
one expects that these excitations give the most important contribution to the
long-wave axial fluctuations of the phase. 
The solution of the hydrodynamic counterpart of the Bogoliubov-de Gennes
equations in the Thomas-Fermi limit for such low-energy axial 
modes yields the spectrum 
$\epsilon_j=\hbar\omega_x\sqrt{j(j+3)/4}$~\cite{Stringari},
where $j$ is a positive integer. The wavefunctions
$f_j^+$ of these modes have the form
\begin{equation} 
\label{fpm} 
f_j^{+}({\bf r})=\sqrt{\frac{(j+2)(2j+3)gn_0({\bf r})}
{4\pi (j+1)R_{\mbox{\tiny TF}}^2L_{\mbox{\tiny TF}}\epsilon_j}}
P_j^{(1,1)}\left(\frac{x}{L_{\mbox{\tiny TF}}}\right),  
\label{trzy}\end{equation}
where $P_j^{(1,1)}$ are Jacobi polynomials. 
Note that the specific form of the
excitation spectrum and the $x$-de\-pen\-dence of the mode functions results
in the quasi 1D limit from the integration of the mode functions over the
transverse directions, as pointed out for instance in Ref.~\cite{Stringari}.

The thermal fluctuations of the phase behave for distances shorter than
$\simeq 0.4 \, L_{\mbox{\tiny TF}}$ as:
\begin{equation}
\label{center}
\langle[\delta\hat\phi(x,x')]^2\rangle_T=\delta_L^2|x-x'|/L_{\mbox{\tiny TF}},
\end{equation}
where $\delta_L^2$ is a measure for the amount of phase fluctuations present in the condensate
and given by~\cite{Shlyapnikov3D}
\begin{equation}
\label{deltaL}
\delta_L^2(T)=T/T_\phi=32\mu k_B T/15N_0(\hbar\omega_x)^2, 
\end{equation}
where $k_{\mbox{\tiny B}}$ is the Boltzmann constant
and $l_{\phi}=L_{\mbox{\tiny TF}}/\delta_L^2$ can be interpreted as a phase coherence length.
Substituting the chemical potential in the
Thomas-Fermi regime in Eq.~(\ref{deltaL}), one obtains the direct dependence on the 
experimental parameters:
\begin{equation}
\!\!\!L_{\mbox{\tiny TF}}/l_{\phi}=\!16\left(\frac{a m^{1/2}}{15^{3/2}
\hbar^3}\right)^{2/5}\!\! \frac{k_{\mbox{\tiny B}}T}{N_0^{3/5}}\left(\frac{\lambda}{\omega_x}\right)^{4/5}.\!\!
\label{deltaLexp}
\end{equation}
Strong phase fluctuations are thus associated with high trap aspect
ratios $\lambda=\omega_\rho / \omega_x$, weak axial confinement $\omega_x$, high
temperatures $T$ and small numbers of condensed atom $N_0$.

The signature of a fluctuating phase can be observed experimentally 
as density modulations (stripes) in the ballistic expansion.
The formation of these stripes can be understood as follows:
After switching 
off the trap, the mean field
interaction rapidly decreases and the axial phase pattern
is converted
into a velocity distribution.
This spatial velocity distribution leads to the appearance
of stripes in the density after a time-of-flight.
These stripes have been described 
analytically~\cite{Hellweg} using a generalization of the Bogoliubov-de Gennes
equation for the self-similar solution
(see e.g.~\cite{Dalfovo,KaganP,castin_dum}).
This equation describes 
the ballistic expansion in presence of fluctuations.
By denoting the density as $n_0+\delta n$ and the phase as
$\phi_0+\phi$
we obtain the analytic expression 
for the relative density fluctuations
\begin{equation}
\frac{\delta \hat n}{n_0}=2\sum_j \tau^{-(\epsilon_j/\hbar\omega_\rho)^2}
\sin \left[\frac{\epsilon_j^2\tau}
{\mu\hbar\omega_\rho (1-(\frac{x}{L_{\mbox{\tiny TF}}})^2)}\right]\hat\phi_j,
\label{delta}
\end{equation}
where the sum extends over the axial modes,
$\tau=\omega_\rho t$, $t$ is the time-of-flight 
and $\hat\phi_j$ is the contribution of the $j$-th 
mode to the phase operator in Eq.~(\ref{operphi}).

By taking the thermal average of the square of Eq. (\ref{delta})~\cite{Hellweg}, 
one obtains a closed relation for the mean square density fluctuations~\cite{square_added}

\begin{equation}
\label{mean_square_fluctuations}
\left \langle \left (\frac{\delta n (x,t)}{n_0(x)} \right )^2 \right \rangle =
\frac{T}{T_\phi} C^2(N_0,\omega_\rho, \omega_x, x,t),
\end{equation}
where
\begin{eqnarray}
&& C^2(N_0,\omega_\rho, \omega_x, x,t)= \nonumber \\
&& \frac{1}{2} \sum_{j=1}^{\infty}\sin^2 \left(\frac{(j+3/2)^2}
{4\alpha(1-(x/L_{\mbox{\tiny TF}})^2)}\right) e^{-\left(\frac{\omega_x} 
{\omega_\rho}\right)^2 \frac{(j+3/2)^2}{2}\ln(2\omega_\rho
t)} \times
\nonumber \\
&&  
\left ( \frac{(j+2)(2j+3)}{j(j+1)(j+3)} \right)
\left(P_j^{(1,1)}\left(\frac{x}{L_{\mbox{\tiny TF}}}\right)\right)^2,
\label{Ccoef}
\end{eqnarray}
with $\alpha=\mu/\hbar\omega_x^2t$.

\section{Experimental Results}
\label{experimental}

\subsection{Experimental setup}\label{subsection_setup}

The experimental setup used to observe phase-fluc\-tu\-a\-ting
Bose-Einstein condensates has been described elsewhere~\cite{Dettmer,Hellweg}.
Here we emphasize the main
features of the apparatus and the technical changes that have led
to the observation of more pronounced effects of phase
fluctuations in the ballistic expansion of an elongated
condensate.

We have adapted the trap geometry to obtain aspect ratios $\lambda>100$
and reduced axial confinement. These high trap aspect ratios are
achieved by axial decompression of our cloverleaf 
trap~\cite{cloverleaf}, lowering the currents in both the pinch
coils and their bias coils. 

For the experiments reported here, a Bose-Einstein condensate of
$^{87}$Rb atoms in the $|F\!=\!1, m_{F}\!=\!-1\rangle$ state is
produced in the axially decompressed trap with 
axial and radial trapping
frequencies 3.4~Hz and 380~Hz 
respectively, by using the following
procedure. We load a magneto-optical trap with a few times $10^8$
atoms from a chirp slowed thermal beam, followed by a short period
of sub-Doppler cooling. Part of the sub-Doppler cooling stage
operates without repumping the atoms from the $|F\!=\!1\rangle$
state, letting them accumulate in the lower hyperfine state. After
optical pumping to the low field seeking $m_F=-1$ state, the atomic
cloud is loaded into a cloverleaf magnetic trap. A cleaning light
pulse on resonance with the $|F\!=\!2\rangle \rightarrow
|F'\!=\!2\rangle$ transition then removes all remaining
$|F\!=\!2\rangle$ atoms from the trap. Finally, the atomic cloud is
adiabatically compressed to allow efficient radio frequency
evaporative cooling. The condensate is loaded in the axially
decompressed trap by using a two step evaporation procedure. The
initial evaporative cooling is performed in our standard trap with
aspect ratio 26, then the trap is ramped to the high aspect ratio
configuration, and the final evaporation is carried out in that
trapping geometry. 
After waiting for 4~seconds to allow the system
to reach an equilibrium state we switch off the trapping potential
within $200\, \mu $s and wait for a variable time-of-flight before
detecting the atomic cloud by resonant absorption imaging.

The experimental observation of phase fluctuations profits from
this adapted trap configuration as indicated by Eq.~(\ref{deltaLexp}).
Our present
measurements show strong fluctuations both due to the high aspect
ratio and the weak axial confinement of our trap.
These lead to a large axial condensate size
(in the elongated trap, $2L \sim 300\,\mu$m) with respect to
the detection resolution of $\sim 5\, \mu$m,
thus favoring the observation of structures in the condensate density after
time-of-flight.

\subsection{Expansion dynamics}
\label{balexpansion}

Phase fluctuations transform into
density modulations during ballistic expansion.
It was shown that the general dependence
on temperature, trapping potential and number of atoms given in
Eq. (3) of~\cite{Dettmer} agrees qualitatively with the
experimental observations. Most data was acquired after a long
time-of-flight of 25~ms since the effect of phase fluctuations is
best observed then. However, here we present observations of the
dynamical transformation of phase fluctuations into density
modulations and quantitatively compare them to the theoretical predictions.

\begin{figure}
\resizebox{0.37\textwidth}{!}{%
  \includegraphics{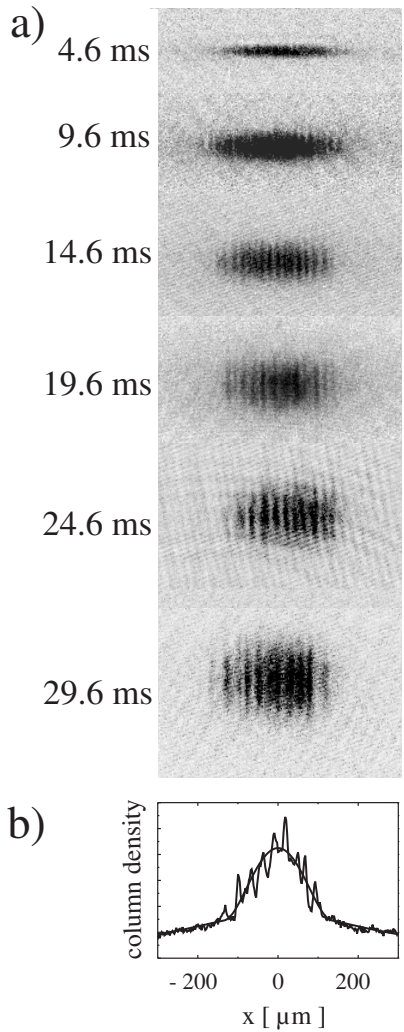}
} \caption{(a) Absorption images of the expanded clouds after various
times-of-flight. The images show $ \sim 7\times 10^4$ condensed atoms at a
temperature of $\sim 0.6 \ T_{\mbox{c}}$. The atomic clouds
are released from a magnetic trap with axial and radial
frequencies of 2$\pi \times 3.4$~Hz and 2$\pi \times 380$~Hz respectively. Due
to the destructive imaging technique, each image shows a different condensate.
(b) The line density profile of the last time-of-flight image is compared with
the corresponding bimodal fitting function.} \label{fig_exp1}
\end{figure}

Fig.~\ref{fig_exp1}(a) shows typical images of the ballistically
expanded clouds for various times-of-flight $t$. The atomic
samples were released from a magnetic trap with axial and radial
trapping frequencies of 3.4~Hz and
380~Hz, respectively. They contain $ \sim 7 \times 10^4$ condensed atoms at a
temperature of $ \sim 0.6 \ T_{\mbox{c}}$. Under these experimental conditions, 
the phase coherence length
in units of the condensate size is
$l_{ \phi } /2L \sim 1/12$.
The usual anisotropic expansion of the
condensates is clearly visible. The line density profile in
Fig.~\ref{fig_exp1}(b) shows the large influence of
phase fluctuations on the expansion of the condensate.
To avoid high optical densities, which
would influence the visibility of density modulations, we 
detuned the detection laser with respect to the atomic resonance
for short times-of-flight.

The amount of phase fluctuations present in a given BEC was determined by
comparing the observed line density profile with the expected bimodal profile
when no phase fluctuations are present. In this bimodal profile the BEC
fraction is described by a parabolic Thomas-Fermi component and the thermal
cloud by  a Gaussian distribution, both integrated along the radial
direction. The deviation of the
experimental density profile from the fitted function is calculated at every
data point in the interval $-0.5\!<\!x/L\!<\!0.5$ and normalized by the fitted 
condensate density at that
point. The average of the squares 
of these values is called $\sigma^2_{\mbox{\tiny BEC}}$ and
represents a normalized spatial average of the
phase fluctuations in a single condensate. The temperature and particle number
of each condensate were determined by separate 2D fits to the absorption
images.

\begin{figure}
\resizebox{0.48\textwidth}{!}{%
  \includegraphics{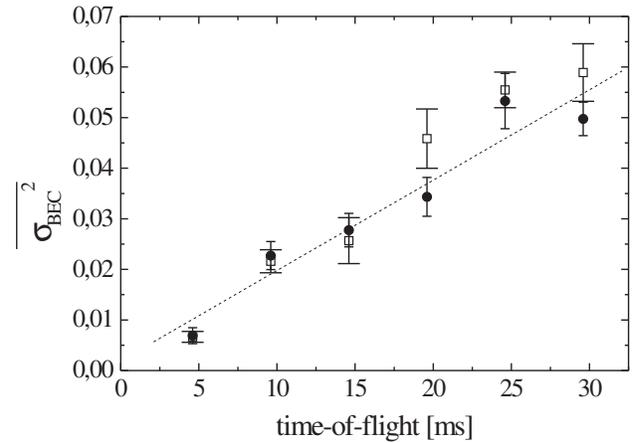}
} \caption{Black circles: Measurements of the normalized spatial average 
of the phase fluctuations $ \overline{\sigma_{\mbox{\tiny BEC}}^2}$ as a
function of the time-of-flight. Each data point is the average of
15 measurements. Open squares: Theoretically predicted values.
Line: Fit to the measured data as a guide to the eye.
The error bars indicate the statistical error.}
\label{fig_exp2}
\end{figure}

Note that phase fluctuations in an elongated BEC are
stochastic. 
During the expansion, the instantaneous phase of the BEC
at the time of release is converted into density modulations and
therefore images taken under the same initial conditions can look
significantly different. Indeed, we observe a large spread of our
experimental data and, therefore, each data point in
Fig.~\ref{fig_exp2} represents the average of 15 measurements.

Fig.~\ref{fig_exp2} shows the observed density modulations
(black circles) as a
function of the free expansion time. All images were taken under
the conditions given in Fig.~\ref{fig_exp1}. The data clearly
shows how the effect of phase fluctuations on the expansion
dynamics increases with the time-of-flight.

To compare our measurements with theory, each
experimental realization (each condensate)
was modeled individually as explained below
and for each time-of-flight the average of all experimental
realizations is compared to the average of all modeled
realizations.
Using the experimentally determined
temperature and atom number, the 
expectation value of the density modulations for each realization was
calculated according to Eq.~(\ref{mean_square_fluctuations}).
Analogous to the experimental procedure, 
a normalized spatial average over the range $-0.5\!<\!x/L\!<\!0.5$ was deduced. 
The theoretical prediction in Fig.~\ref{fig_exp2}
takes the limited resolution of our imaging system
into account~\cite{Hellweg}.

We obtain very good quantitative agreement between our results and the
theoretical predictions, confirming that the transfer mechanism of
phase fluctuations into density modulations during time-of-flight
is well understood.

\subsection{Suppression of density fluctuations}
\label{density}

The theoretical description of phase fluctuations 
is based on the prediction~\cite{Shlyapnikov1D} that
density modulations are strong\-ly suppressed in the magnetic trap.
In the Thomas-Fer\-mi
regime, where the mean field energy dominates over the kinetic
energy, the excitation of density modulations requires a high
energetic cost
on the order of the chemical potential.
The measurements reported in Fig.~\ref{fig_exp2} clearly show that the 
observed density modulations are suppressed as the time-of-flight
approaches zero. However, this does not rule out the existence of 
density modulations on a length scale smaller than our 
experimental resolution.

The suppression of density fluctuations can be verified
experimentally by measuring the second order correlation function
$g^{(2)}({\bf r}_1-{\bf r}_2)$ of the field operator $\hat{\Psi}({\bf r})$. In
second quantization formalism,
\begin{equation}
g^{(2)}({\bf r}_1-{\bf r}_2)=\frac{\langle\hat{\Psi}^{\dagger}({\bf r}_1)
\hat{\Psi}^{\dagger}({\bf r}_2)
\hat{\Psi}({\bf r}_2)\hat{\Psi}({\bf r}_1)\rangle}{n({\bf r}_1)n({\bf r}_2)}.
\label{g2}
\end{equation}
In particular, $g^{(2)}(0)$ gives the correlation
function of the atomic density for ${\bf r}_1 = {\bf r}_2$.
As shown in~\cite{Ketterle},
$g^{(2)}(0)$ is directly related to the expectation value of the
interaction energy $U$ by

\begin{equation}
\!\!\langle U \rangle=\frac{2 \pi \hbar^2
a}{m}g^{(2)}(0)\int{\mbox{d}^3r\, n^2({\bf r})}
.\!\!
\label{U}
\end{equation}
In the Thomas-Fermi regime, only the interaction energy $\langle U
\rangle$ contributes to the kinetic energy of the condensate after
ballistic expansion. 
The presence of density fluctuations
would however lead to an increase of the observed
release energy due to the presence of
repulsive
particle interactions. Hence, the ratio
of the release energy and the calculated interaction
energy in Thomas-Fermi approximation
gives the value of $g^{(2)}(0)$. The
energy due to phase fluctuations is small
compared to the interaction energy (for typical parameters
$\sim 0.5\%$) and can therefore be neglected in release
energy measurements.

From earlier release-energy measurements, the product $a \, g^{(2)}(0)$
has been determined~\cite{g2_jila,castin_dum,cloverleaf}.
We have performed such a measurement for phase-fluctuating condensates.

\begin{figure}
\resizebox{0.48\textwidth}{!}{%
  \includegraphics{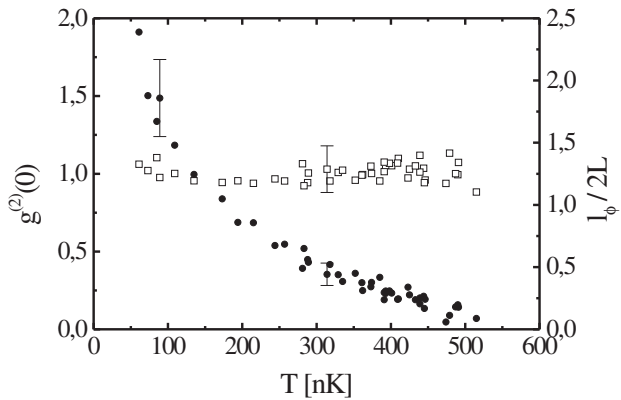}
} \caption{Open squares: Measurements of $g^{(2)}(0)$ as a
function of temperature for phase-fluctuating $|F\!=\!1\rangle$
condensates in a $\lambda=30$ trap. Black dots: Calculated
phase coherence length $l_\phi$, in units of the axial condensate
size $2L$, for the experimental conditions of each data point. The error bars
indicate the maximum error.}
\label{fig_exp3}
\end{figure}

Fig.~\ref{fig_exp3} shows $g^{(2)}(0)$ (open squares) as a function of
temperature for atoms in the $|F\!=\!1\rangle$ state in a trap
with $\lambda=30$. We use this weakly elongated trap, 
because this trapping geometry enables us to use the sample 
temperature to tune from nearly pure to strongly 
phase-fluctuating condensates experimentally.

The
same graph also shows the calculated phase coherence length
$l_\phi$ according to Equation~(\ref{deltaLexp}) for the experimental
conditions of each data point (black circles). As the temperature of 
the sample is raised, the phase coherence length decreases
and becomes significantly shor\-ter than the condensate size.
Thus, our measurements clearly reach the regime of quasicondensation.
However, the graph shows that the
second-order correlation function is largely independent of the
condensate temperature $T$, and that it is consistent with the
expected value $g^{(2)}(0)=1$. These measurements give an upper
limit for the density modulations present in a phase-fluctuating
condensate, and clearly distinguish our samples from thermal clouds
where, due to bunching effects, $g^{(2)}(0)=2$ and 
large density fluctuations are present on small scales.

\section{Controlling the phase}
\label{subsection_theory1}
Parametric resonance refers to the exponentially large response
of a system of a periodic external 
perturbation~\cite{param}
for some specific set of parameters.
Although parametric resonance is a very well established phe\-no\-me\-non
in linear systems, it was shown in~\cite{garciari1}
that parametric resonance can also occur in Bose-Einstein condensates. 
Normally one aims at using parametric resonance
in order to modulate the condensate density 
by means of a relatively small periodic perturbation. 
However, in the Thomas-Fermi regime, as mentioned above,
modulations of the density require large energies.
In this Section we suggest a way to shape the overall phase
by modulating the trap with
a small amplitude for a short time.

\subsection{Theoretical methods}

To illustrate the effect of the perturbation on 
the phase of a quasicondensate,
we numerically solve the 1D Gross-Pitaevskii equation for different initial
conditions. We assume a zero temperature BEC in the Thomas-Fermi regime
containing $5\times 10^4$ $^{87}$Rb atoms 
trapped in a 1D confining potential with an (axial) frequency
of $\omega_{x}=2 \pi\times 14$ Hz.
The parabolic density profile is then given by
$n_0(x)=n_{0\mbox{m}}(1-x^2/L^2_{\mbox{\tiny{TF}}})$
where $L_{\mbox{\tiny TF}}=(2\mu/m\omega^2_x)^{1/2}$ and $n_{0\mbox{m}}=\mu/g$. 
The coupling constant $g$ in one dimension can be derived 
by averaging the 3D interactions over the radial density profile.
We first evolve the wavefunction according to the
time dependent GP equation in imaginary time
to obtain the condensate
wavefunction at $T=0$. We ensure that the initial phase of the condensate
is constant before imposing a fluctuating 
phase corresponding to a fixed temperature $T$
on the condensate wavefunction.
In strictly 1D, one expects that for temperatures 
$T_{\mbox{\scriptsize d}}\gg T \gg T_{\phi}$ 
phase fluctuations are present whereas density fluctuations are 
suppressed~\cite{Shlyapnikov1D}. Here
$k_{\mbox{\tiny B}} T_{\mbox{\scriptsize d}}\approx N\hbar\omega_x$ is 
the degeneracy temperature~\cite{ketterle1d},
 and $T_{\phi}=T_{\mbox{\scriptsize d}}\hbar\omega_x/\mu$ 
is the characteristic temperature for
the appearance of the phase fluctuations.

Applying the same procedures for the 1D case as
described in Section 2 for
the 3D case, 
the field annihilation operator can be written as
$\hat \psi(x,t=0)=$
$\sqrt{n_0(x)}\exp(i\hat\phi(x))$. The
operator of the phase is given by 
(see~\cite{Shlyapnikov1D,Shev})
\begin{equation}
\label{operphi2}
\hat\phi({x})=\frac{1}{\sqrt{4n_0(x})}\sum_{j=1}^{\infty}
f_j^{+}({x})\hat a_j +\mbox{h.c.}
\end{equation}
where, as before, $\hat{a}_j$ is the annihilation operator of the excitation with
quantum number $j$, $f_j^{+}= u_j +
v_j$ and $u_j$, $v_j$ are the excitation functions
determined through the
Bogoliubov-de Gennes 
equations in the Thomas-Fermi limit.
The solution of these 1D equations gives the spectrum
$\epsilon_j=\hbar\omega_x\sqrt{j(j+1)/2}$~\cite{1despec}.
The functions $f_j^{+}$ have now the form
\begin{equation}
\label{fpm1d} 
f_j^{+}(x)=\sqrt{\frac{(j+1/2)}{L_{\mbox{\tiny{TF}}}}\left[\frac{2\mu}{\epsilon_j}
(1-z^2)\right]}P_j(z),
\end{equation}
where $P_j(z)$ are the Legendre polynomials and
$z=x/L_{\mbox{\tiny{TF}}}$. Note, that these solutions correspond to the
strictly 1D case and differ from those obtained in the quasi 1D case
described by (\ref{trzy}). The qualitative and, to great
extent, quantitative character of the solutions (\ref{fpm1d}) and
(\ref{trzy})
is, however, similar.
  
The random phase of the condensate is numerically simulated 
by replacing the operators
$\hat a_j$
and $\hat a^{\dagger}_j$ in Eq.~(\ref{operphi2}) by Gaussian 
random variables $\alpha_j$
and $\alpha^{*}_j$, with the correlation 
 $\langle \alpha^{*}_j\alpha_{j'} \rangle =\delta_{jj'}N_j$, 
where $N_j$ is the occupation number for the quasiparticle mode 
$j$ for a given temperature $T$.

Once the phase is imposed, we apply a periodic perturbation
$\alpha\sin(\omega_{\mbox{s}} t)$ to the trapping potential:
\begin{equation}
\omega^2_x\to \omega^2_x(1+\alpha \sin(\omega_{\mbox{s}} t)).
\end{equation}
The amplitude $\alpha$ of the
perturbation is small and ranges between $0.05$ and $0.20\ll 1$, 
whereas the frequency $\omega_{\mbox{s}}$ ranges from $\omega_x$ to several times $\omega_x$.
The perturbation acts for a short period of time $t$ 
(the simulations were performed for 2, 6 and 12 periods of modulation $2\pi/\omega_{\mbox{s}}$). 
After switching off the perturbation 
the amplitudes of the different modes participating in 
the fluctuations of the phase are calculated.
The thermal fluctuations of the
phase for the temperature range considered here  
are mostly provided by the low excitation modes ($j< 50$). 
We calculate the amplitude of these modes as a function 
of the frequency of the perturbation.

\subsection{Numerical results}
Our calculations show that the positive and negative frequency modes
couple resonantly and that by appropriately tuning 
the frequency of the perturbation one can selectively suppress some modes
and enhance others. The enhancement, however, is not particularly strong, 
since we do not amplify any mode (except the second mode which arises
directly due to the parametric modulation) 
much above its initial value provided by
Eq.~(\ref{fpm1d}). 
Note that the odd modes can only be excited if the symmetry is broken,
e.g. by the existence of phase fluctuations.

\begin{figure}
\centering{\resizebox{0.37\textwidth}{!}{%
\includegraphics*{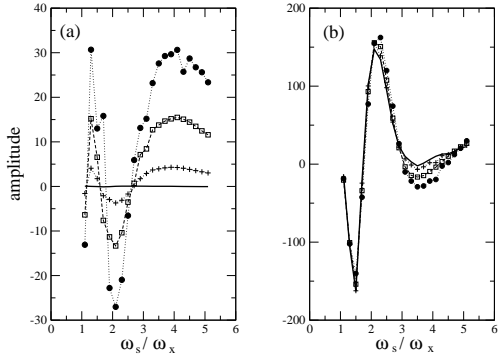}
} \caption{(a) Amplitude of the first and (b) second mode 
           as a function of the modulation frequency $\omega_{\mbox{s}}$.
           The perturbation was applied for $t=2\times 2\pi/\omega_{\mbox{s}}$
           with an amplitude of $\alpha=0.05$. 
The solid line indicates the final amplitude starting 
from a pure condensate ($T=0$); crosses correspond to a quasicondensate
at initial temperature of $T=0.01 \, T_{\mbox{c}}$, open squares to $T=0.1 \, T_{\mbox{c}}$ 
and circles to $T=0.4 \, T_{\mbox{c}}$. The same set of random coefficients were used in all the cases.
}
\label{fig_the1}}
\end{figure}

Figure~\ref{fig_the1}
displays the amplitudes
of the first two modes for different initial
temperatures as a function of the frequency $\omega_{\mbox{s}}$. 
The second mode, which corresponds
to the density breathing mode, 
is strongly enhanced due to the perturbation of the
trap. This mode amplitude can be calculated in the Thomas-Fermi
regime by using a self similar solution 
of the GP equation~\cite{KaganP,castin_dum} describing
the dynamics of the bare condensate that 
implies appearance of a phase quadratic in $x$.
Indeed, only for sufficiently large temperatures ($T\sim 0.5 \, T_{\mbox{c}}$), 
its frequency dependence starts to deviate significantly from
its behavior at $T=0$.

\begin{figure}
\centering{\resizebox{0.4\textwidth}{!}{%
 \includegraphics*{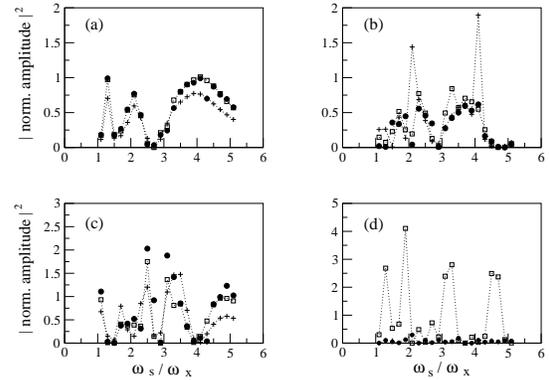}
} \caption{Modulus square of the exciting mode 
           amplitudes normalized to their initial
           value at $t=0$. The contribution corresponding
	   to $T=0$ has been previously 
           subtracted. (a) (b) (c) and (d) correspond to amplitudes 
           of the first second, third and fourth mode
           for the same parameters as Fig. \ref{fig_the1}.}
\label{fig_the2}}
\end{figure}

To show the effect of the fluctuating phase on
the final amplitudes, we subtract
the phase a pure condensate
would have acquired under the same perturbation 
from our results.
Figure~\ref{fig_the2} displays the modulus square of such amplitudes
normalized to their initial values for the first four modes.
The amplitudes of the high excited modes (app. $j > 4$) are
practically equal and neglibigly small compared to the lower modes
therefore the lower modes
dominate the dynamics of the phase. 
Only those modes show clear 
maxima and minima as a function of the modulation frequency
$\omega_{\mbox{s}}$.
Resonances between different modes may appear when 
\begin{equation}
\omega_{\mbox{s}}=|\Omega_i\pm\Omega_j|,
\end{equation}
where $\Omega_i=\epsilon_i/\hbar$. In order to observe resonances 
it is necessary that the energy differences between the modes 
correspond to multiples of the modulation frequency,
otherwise the system behaves irregularly.
In the 1D case, the excitation modes are quite regularly separated 
by approximately $\omega_x/\sqrt{2}$ ($\approx 0.71\, \omega_x$).
However, in nonlinear
systems, the resonance frequencies undergo shifts which can already be 
quite significant at small perturbations~\cite{burnett}.
For the first mode (Fig.~\ref{fig_the2}), we find clear 
resonances at $\omega_{\mbox{s}}/\omega_x \simeq 1.4$, $2.1$ and $4.0$. One is tempted to attribute
the first two resonances to the coupling with $j=3$ and $j=4$ mode.
The second mode, which dominates the dynamics --- since this mode
is enhanced by the perturbation --- shows three maxima at
$\omega_{\mbox{s}}/\omega_x\simeq 1.4$, $2.1$ and $3.6$.
The analysis of maxima for 
the third and higher modes becomes quite complex. 
For the third mode, we find maxima 
at $\omega_{\mbox{s}}/\omega_x\simeq 1.4$, $1.7$, $2.4$, $3.4$ and $5.0$.
Starting from the fourth mode it becomes difficult to
associate the maxima
with specific resonances.

\begin{figure}
\centering{\resizebox{0.3\textwidth}{!}{%
 \includegraphics*{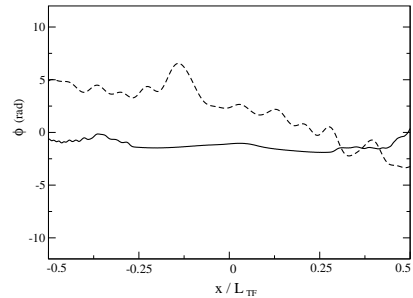}
} \caption{The dotted line shows a typical phase pattern 
for $\omega_{\mbox{s}}=2\pi \times 14$ Hz,
$\lambda=51$ and initial temperature $T=0.4 \, T_{\mbox{c}}$,
the solid line shows the final phase after applying 
the time dependent perturbation for a duration of
two cycles with $\alpha=0.05$ and
$\omega_{\mbox{s}}=5.7 \, \omega_x$.}
\label{fig_the3}}
\end{figure}

It is important to stress here that for the lower modes, the position 
of the maxima does not depend
on the random generated phase.
This fact clearly confirms that we are dealing here with resonances 
between different modes. 
By increasing the perturbation time the spectrum changes,
as expected,
since this is a time dependent effect,
and more peaks appear.
However with the exception of the second one, no enhancement
of the modes is observed.
By increasing the amplitude $\alpha$ of the perturbation, 
the second mode becomes much more enhanced, 
but again  the amplitudes of the other modes
(normalized to their initial values at $t=0$) do not.
In other words, we
do not find an exponential parametric amplification by tuning the
frequency to $2\Omega_i/n$ where
$n$ is a positive integer. Nevertheless, the fact that the amplitudes
of the lower modes display well defined maxima and minima
allows to shape the overall phase of the quasicondensate. 
As an example, in Fig.~\ref{fig_the3} we show
the initial fluctuating phase 
of a quasicondensate at $T=0.4 \, T_{\mbox{c}}$, and
its final phase after the perturbation
is switched off. In the range $|x/L_{\mbox{\tiny TF}}|<0.5$ the final phase 
becomes practically constant 
demonstrating that the initial fluctuations
can be significantly suppressed using the parametric modulation technique.
To achieve this specific phase we chose the frequency $\omega_{\mbox{s}}$
such that the lower modes have approximately the same amplitude value.

\section{Conclusion and outlook} \label{subsection_conclusion}

In this paper we analyze fluctuations of the phase of quasi 1D condensates
at temperatures of the order of fractions of
$T_{\mbox{c}}$.
The dynamical transformation of phase fluctuations into
density modulations was observed as
function of free expansion time. A detailed comparison
of the statistical average of the modulation
with the theoretical prediction shows excellent
agreement. It was also verified experimentally 
that density modulations of a phase fluctuating
BEC in the trap are strongly suppressed. The
quantitative understanding of the transfer of phase fluctuations
into density modulations opens a pathway to
use phase fluctuations for condensate thermometry.

Phase fluctuations impose restrictions 
on the applicability of quasi 1D condensates in atom optics and 
precision atom interferometry.
%One obvious way to reduce the phase fluctuations is to reduce
%the temperature of the BEC. 
We show that a possible mechanism to control
the overall phase of a 
quasi 1D condensate is to use a small periodic perturbation of the
trap. In this way, only few of the modes responsible for the phase 
fluctuations become relevant. Furthermore, by properly tuning
the frequency, time and amplitude of the external perturbation
one can resonantly couple the different relevant low order modes, 
so that some of them are enhanced and others are inhibited.
This method suggest a pathway to control phase fluctuations 
and overcome their undesired effects.

We acknowledge valuable discussions with G. Shlyapnikov, D. S. Petrov
and C. Bord{\'e}.
This work is supported by the \textit{Deutsche Forschungsgemeinschaft}
within the SFB\,407 and the Schwerpunktprogramm "Wechselwirkungen ultrakalter atomarer
und molekularer Gase", Alexander von Humboldt Stiftung, and
the European Science Foundation (ESF) within the BEC2000+ programme.

%% FURTHER ACKNOWLEDGEMENTS WILL BE INSERTED AFTER THE REVIEW PROCESS


\begin{thebibliography}{}
\bibitem{square_added} Note that Eq.~(\ref{mean_square_fluctuations}) 
was misprinted in~\cite{Hellweg} and that the correct equation is given here.
\end{thebibliography}
\end{document}